\documentclass[12pt,preprint]{aastex}
\usepackage{amsmath}
\usepackage{amssymb}
\usepackage{amsthm}
\usepackage{graphicx}
\usepackage{epstopdf}
\usepackage{url}
\usepackage{algorithmic} 
\usepackage{algorithm}
\usepackage{verbatim}
\usepackage{multirow}
\usepackage{wrapfig}
\usepackage{array}
\usepackage{multirow}

\usepackage{color,psfrag}

\usepackage[titles]{tocloft}

\theoremstyle{definition}

\DeclareMathOperator*{\minimize}{minimize \quad}
\DeclareMathOperator*{\subto}{subject\, to \quad}

\newcommand{\reals}{\ensuremath{\mathbb{R}}}

\newcommand{\norm}[1]{ \Vert #1 \Vert }

\newcommand{\convOp}{\mathcal{C}}

\newcommand{\bSeries}{\boldsymbol{\beta}}

\newcommand{\Err}{\varepsilon}

\newcommand{\vst}{\ensuremath{\text{vst}}}
\newcommand{\eqrefa}[1]{Eq.~\eqref{#1}}
\newcommand{\Ang}{\AA\ }

\usepackage{graphicx}
\usepackage{dcolumn}
\usepackage{bm}

\begin{document}

\title{The First Stray Light Corrected EUV Images of Solar Coronal Holes}

\author{Paul Shearer}
 \email{shearerp@umich.edu}

\author{Richard A. Frazin}%
 \email{rfrazin@umich.edu}
 
 \author{Alfred O. Hero III}
\email{hero@umich.edu}
 
 \author{Anna C. Gilbert}
\email{annacg@umich.edu}

\begin{abstract} 
Coronal holes are the source regions of the fast solar wind, which
fills most of the solar system volume near the cycle minimum. Removing
stray light from extreme ultraviolet (EUV) images of the Sun's
corona is of high astrophysical importance, as it is required to make
meaningful determinations of temperatures and densities of coronal
holes.  EUV images tend to be dominated by the component of the stray light
due to the long-range scatter caused by microroughness of telescope
mirror surfaces, and this component has proven very difficult to
measure in pre-flight characterization.  In-flight characterization
heretofore has proven elusive due to the fact that the detected image
is simultaneously nonlinear in two unknown functions: the stray light pattern
and the true image which would be seen by an ideal telescope.  Using
a constrained blind deconvolution technique that takes advantage of
known zeros in the true image provided by a fortuitous lunar transit,
we have removed the stray light from solar images seen by the EUVI instrument
on STEREO-B in all four filter bands (171, 195, 284,
and 304 \AA).  Uncertainty measures of the stray light corrected images, which
include the systematic error due to misestimation of the scatter, are
provided.  It is shown that in EUVI, stray light contributes up to 70\% of the
emission in coronal holes seen on the solar disk, which has dramatic
consequences for diagnostics of temperature and density and therefore
estimates of key plasma parameters such as the plasma $\beta$\ and ion-electron
collision rates.
\end{abstract}

\maketitle

\section{INTRODUCTION}
One of the longest standing puzzles in astrophysics concerns the processes that energize the solar wind. The fast solar wind, which has a speed of $\sim 800 \text{\, km/s}$ near the Earth, emanates from EUV and X-ray faint regions in the Sun's atmosphere called \emph{coronal holes} \citep{Krieger1973}. Physics-based modeling efforts to identify the processes that heat and accelerate the solar wind have met with very limited success \citep{Cranmer2010}.  It is likely that much more observational input will be required to adequately constrain the numerical models used to investigate this question. NASA's EUV imaging instruments (SOHO/EIT, TRACE, STEREO/EUVI, and SDO/AIA) provide a comprehensive data set covering 1 1/2 solar cycles that can provide powerful constraints on temperatures and densities in coronal holes \citep{Frazin09}. However, coronal holes and other faint structures are severely contaminated with stray light, which must be corrected before these vast data sets may be utilized for this purpose.


All EUV imaging instruments are afflicted to some degree by long-range scattering, which distributes a haze of \emph{stray light} over the whole imaging plane. This haze is not noticeable in the brighter areas of the image, but can completely overwhelm the emissions of faint regions. Stray light arises from two sources: non-specular reflection from microrough mirror surfaces, and diffraction due to the pupil and a mesh obstructing the pupil \citep{Howard08}. Here we describe a method for, and results of, correction of stray light in STEREO-B/EUVI, henceforth abbreviated as EUVI-B.

Stray light correction requires determination of the instrument's point spread function (PSF). The four EUVI channels (171, 195, 284, and 304 \AA) have different optical paths and therefore different PSFs. Except for pixel-scale variations due to optical aberration (whereas the scattering of interest here is significant over ranges of hundreds of pixels), the EUVI PSFs are spatially invariant \citep{Howard08}. Thus, stray light contamination is modeled by convolution with the PSF, and the
correction process is \emph{deconvolution} \citep{Starck02}. 

The instrument PSFs are difficult to characterize experimentally due to the lack of a sufficiently strong EUV source, so they must be determined primarily from in-flight observations. Solar flare images can provide significant information about the entrance aperture diffraction \citep{Gburek:TracePSF171:2006}, as diffraction orders are easily visible around the flare. Unfortunately, most of the scatter is too diffuse to be observed clearly around flares. The best information about diffuse scatter is obtained from transiting bodies that do not emit in the EUV, so any apparent emission is instrumental in origin. DeForest et al. \citep{DeForest:CoronaStrayLightTRACE:2009} used a Venus transit to estimate the mirror scattering in the TRACE instrument by fitting a truncated Lorentzian. Here, we present the first self-consistent determination of the mirror scattering via blind deconvolution, in which both the true solar image and various PSF parameters are taken to be unknown simultaneously.   The information making this effort successful comes from calibration rolls and the Feb.~25, 2007 STEREO-B lunar transit, each exposure of which provided about 50,000 lunar disk pixels illuminated only by instrumental effects.

\section{BLIND DECONVOLUTION METHOD}
We characterize scattering in a given EUVI filter band with a three-component PSF. The first two components, $h^p$ and $h^g$, account for diffraction through the pupil and pupil mesh, and were determined analytically using Fraunhofer diffraction theory \citep{Goodman:FourierOptics}. The third component, $h^m$, accounts for the remaining diffuse scatter, much of which derives from the mirror microroughness. The formula for this component contains free parameters $\varphi$ which we determine empirically by blind deconvolution, and we write $h^m_\varphi$ to indicate the dependence. The total PSF $h_\varphi$ is the convolution of these three components: 
\begin{equation}
h_\varphi = h^g * h^p * h_\varphi^m,
\end{equation}
where
\begin{equation}
(u * v)(x) = \sum_{x' \in I} u(x - x') v(x')
\end{equation}
denotes the convolution of two discrete functions $u$ and $v$ over the index set $I$ of the $2048 \times 2048$ array of CCD pixels. (We set $u(x-x') = 0$ at pixels $x-x'~\notin~I$.) This model assumes that each component is independent, neglecting phase correlations that arise as light propagates from the entrance aperture to the primary mirror. To determine the proportionality constants of the three PSFs, we assume that each sums to unity. 

We used the EUV mirror literature to help us choose an appropriate parametric model for the empirical PSF component $h^m_\varphi$. In a typical EUV scatter model, a significant fraction of the light is not scattered, while the rest is scattered by broad wings \citep{Krautschik:EUVscatter:2002}. Up to scaling and normalization constants, these wings are described by the power spectral density (PSD) of the mirror surface height function, and this PSD has been directly measured for mirrors similar to EUVI's  \citep{Galarce:EUVModeling:2010}. A log-log plot of the measured PSD versus spatial frequency is roughly piecewise linear, implying that $h_\varphi^m$ is a piecewise power law whose exponent depends on pixel distance $r$ from the origin.  

Accordingly, a parametric formula for a family of piecewise power laws was used to describe $h_\varphi^m$. A series of breakpoints $1 = r_0 < r_1 < r_2 < \ldots < r_b < \infty$ was chosen, and the number of breakpoints, $b=8$, was selected according to a $\chi^2$ goodness of fit criterion (we performed several fits using different values of $b$, and the fit did not improve significantly for $b > 8$). On each subinterval $[r_{i-1},r_i)$, the formula is given by $p_{\alpha,\bSeries}(r) \propto r^{-\beta_i}$, where $\beta_i \geq 0$, and $\bSeries = (\beta_1,\ldots,\beta_b)$. The parameter $\alpha$ represents the fraction of non-scattered light: $p_{\alpha,\bSeries}(0) = \alpha$.  This radially symmetric profile did not adequately describe the anisotropic scatter observed in the transit and calibration roll images, even after removal of the scatter due to $h^p$ and $h^g$. We therefore generalized the formula to allow $h^m_\varphi$ to have an elliptical cross section. Let $M_{s,\theta}$ denote the $2 \times 2$ matrix that dilates the plane by a factor of $s$ along a line rotated $\theta$ radians counterclockwise from the horizontal axis; then

\begin{equation} \label{hmrBuild}
h^m_\varphi(x) = p_{\alpha,\bSeries}(\norm{M_{s,\theta} \cdot x}_2+\delta),
\end{equation}
where the constant $\delta = 1$ was added to avoid the power law's singularity at the origin.  The free parameters of the PSF model $h_\varphi$ are then $\varphi = (\bSeries,\alpha,\theta,s) \in \reals^{p}$, where $p = b+3 = 11$. 

\begin{figure}
\begin{tabular}{c}
\includegraphics[scale=2.00]{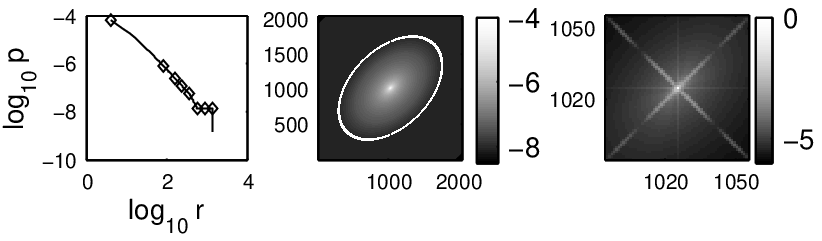} \\
\end{tabular}
\caption{An overview of the 171~\Ang PSF structure. All distances are in pixels. \emph{Left:} log-log plot of power-law profile function $p_{\alpha,\beta}(r)$ with breakpoints marked. \emph{Center:} total PSF $h_\varphi$ with elliptical contour highlighted in white (logarithmic color scale). \emph{Right:} PSF core. The centrally emanating streaks are diffraction orders from $h^p$ and $h^g$.}
\label{psfModel}
\end{figure}

To determine the PSF from the lunar transit data, we described the observation process by a statistical image formation model and sought a maximum likelihood estimate of the PSF under this model. We let $u^{\text{true}}$ denote the ideal image; that is, $u^{\text{true}}(x)$ is the expected solar photon count that would be measured at pixel $x$ by an ideal instrument. Due to scatter and the Poisson photon arrival process, the actual number of photon arrivals is a Poisson random variable with expected value $h^{\text{true}} * u^{\text{true}}$. The difference between the expected and the actual photon count is the photon noise $n_{\text{phot}}$. The observed photon count $f$ deviates from $h^{\text{true}} * u^{\text{true}} + n_{\text{phot}}$ due to CCD dark current and read noise $n_{\text{ccd}}$. Based on histograms of dark images, we find $n_{\text{ccd}}$ is reasonably modeled as a Gaussian white noise process with standard deviation $\sigma_{\text{ccd}} \approx 1$ digital number (DN). The variance must be divided by the photometric gain factor (the recorded DN per incident photon) to obtain the CCD noise level in units of photons. Combining $n_{\text{phot}}$ and $n_{\text{ccd}}$ into a single variable $n$, we obtain the statistical image formation model
\begin{equation} \label{imageFormModelEqn}
f = h^{\text{true}} * u^{\text{true}} + n.
\end{equation}
We would like to find a maximum likelihood estimate of the PSF under the model of \eqrefa{imageFormModelEqn}, but the Poisson-Gaussian distribution of $n$ leads to a difficult large-scale nonlinear optimization problem. We therefore apply a variance stabilizing transform (VST) to make $n$ approximately standard normal (mean zero and variance one), which leads to an easier least-squares problem. A VST for a Poisson-Gaussian random variable where the Gaussian has variance $\sigma^2$ is derived in \citep{MurtaghStarck:ImRestVST:1995}. Their full formula allows for CCD bias and non-unity camera gain; however, we correct for these in pre-processing, so the formula simplifies to
$\text{vst}(I) = 2\sqrt{I + \frac{3}{8} + \sigma^2}.$
If the VST is applied to both sides of \eqrefa{imageFormModelEqn} with $\sigma^2 = \sigma_{\text{ccd}}^2$, we obtain 
\begin{equation}
\text{vst}(f) \approx \text{vst}(h^{\text{true}} * u^{\text{true}}) + n_{\text{vst}},
\end{equation}
where, for each $x$, $n_{\text{vst}}(x)$ is roughly standard normal \citep{MurtaghStarck:ImRestVST:1995}. The approximation begins to break down when $I \lesssim 5$ photons, but this only occurs far from the solar disk in the lunar transit images. In these low-intensity regions we replace the observed image pixel value with a local average of sufficient size that the total photon counts exceed 5 in the averaged neighborhood.

We found an estimate $h_\varphi$ of the EUVI-B PSF $h^{\text{true}}$ by solving a blind deconvolution problem on a series of 8 images $f_1,\ldots,f_8$ from the lunar transit series. We assume the scatter-free images $u_i^{\text{true}}$ are positive everywhere and zero on the lunar disk pixels $Z_i$, leading to the constraints
\begin{equation} \label{imageConstraints}
u_i^{\text{true}} \geq 0, \quad u_i^{\text{true}}(Z_i) = 0, \quad i=1,\ldots,8.
\end{equation}
The lunar disk was identified by detecting its edge pixels with gradient thresholding, then fitting a circle to the detected edge pixels. An approximate maximum likelihood estimate for the PSF $h_\varphi$ and images $\{u_i\}$ is obtained by solving
\begin{equation} \label{uPhiOptEqn}
\begin{split} 
\minimize_{\varphi = (\bSeries,\alpha,s,\theta), \, \, \{u_i\} \, \, \, \, \, \, \,} 	
&\sum_{i=1}^8 \norm{ \vst(h_\varphi * u_i) - \vst(f_i) }^2 \\				
\subto \, \, \, \, \, \,
&u_i \geq 0, \, \, u_i(Z_i) = 0 \, \, \text{for all $i$,} \\
&\bSeries \geq 0, \, \, 0 \leq \alpha \leq 1.
\end{split}
\end{equation}
Since each $u_i$ is a $2048 \times 2048$ image, this problem has over 32 million variables and is intractable by general-purpose numerical methods. We solved this problem with a customized variant of the variable projection method \citep{Golub:separablenonlinear:2002}, which is efficient for problems of this kind.


\begin{figure}
\includegraphics[scale=1.3]{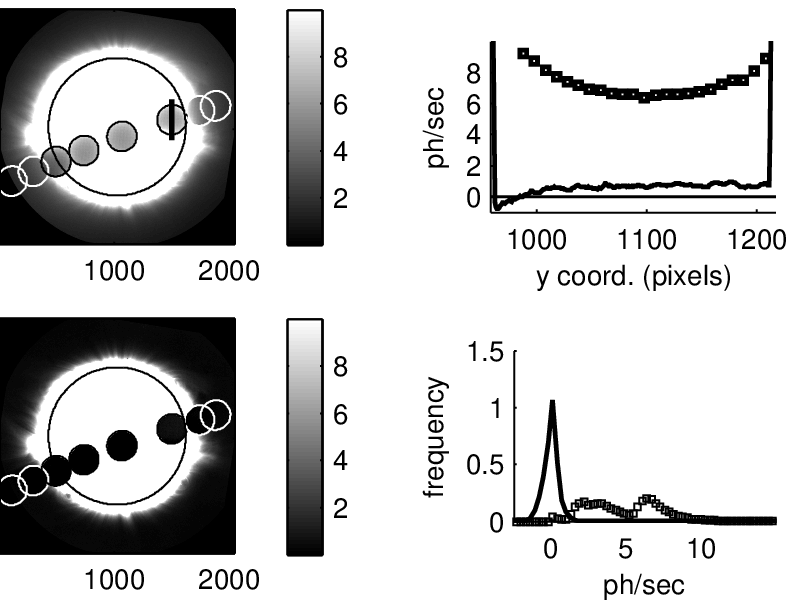} \\
\caption{The lunar transit before and after stray light correction (171~\AA). The lunar and solar limbs are outlined for reference. \emph{Top left:} The lunar disks from the eight transit images $f_i$ superimposed on the first image of the series, $f_1$.  The colorbar is in photons per second (ph/sec) and has a low upper limit to show the stray light on the lunar disks. The black line segment identifies a series of pixels whose intensity values are plotted to the right. \emph{Bottom left:} The lunar disks from the CV images $u_i^\star$ superimposed on $u_1$. \emph{Top right:} Intensity along the vertical black line segment before correction (squares) and after (solid).  \emph{Bottom right:} Normalized histograms giving the distribution of lunar disk intensities before correction (squares) and after (solid). \vspace{-1mm} }
\label{cvAnalysis}
\end{figure}

\begin{figure}
\includegraphics[scale=1.3]{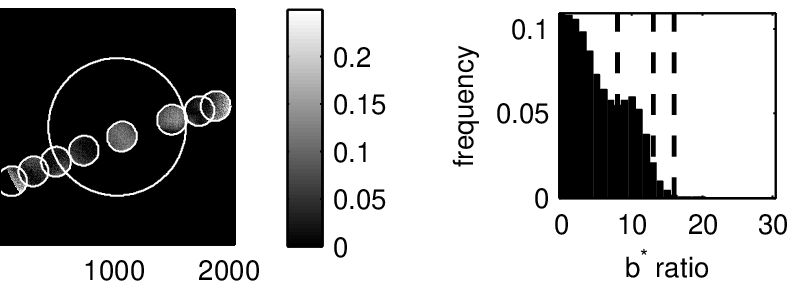} \\
\caption{\emph{Left:} Map of the $b_i^\star$ values on the 8 lunar disks. \emph{Right:} Normalized histogram giving the distribution of the $b_i^\star$ values. The dashed vertical lines indicate the $68^{th}$, $95^{th}$, and $99.7^{th}$ percentile values. \vspace{-1mm} }
\label{cvErrorAnalysis}
\end{figure}

Fig.~\ref{psfModel} shows a log-log plot of the 171~\Ang profile function obtained by blind deconvolution, along with views of the total PSF and its core. Note that the profile is constant for $200 < r < 1200$ pixels, implying considerably more long range scatter than a constant power law exponent would predict. The total PSF is dominated by the elliptical power law decay, although the diffraction orders from the pupil and mesh PSFs are visible near the PSF core.

Once we determined the PSF from blind deconvolution, we were able to correct any observed EUVI-B image $f$ via more standard deconvolution methods. Let $\convOp(h) : \reals^I \rightarrow \reals^I$ denote the linear operator that convolves an input image $u$ with a fixed PSF $h$: $\convOp(h)u~=~h*u$. The stray light correction of a given image $f$ is obtained by applying the inverse of $\convOp(h)$ to $f$:
\begin{equation} \label{uDef}
u = \convOp(h)^{-1}f.
\end{equation}
The inverse operator does not exist for arbitrary $h$. However, our PSFs have an origin value $h(0) > 1/2$, which makes $\convOp(h)$ diagonally dominant and invertible. The inversion of \eqrefa{uDef} can be performed with the conjugate gradient method in less than a minute on a laptop.

\section{MODEL VALIDATION AND ERROR ANALYSIS}
A first test of our stray light correction was performed by applying cross-validation (CV) to the lunar transit series. For each of the 8 images $f_i$, we determined a PSF $h_i$ by solving \eqrefa{uPhiOptEqn} with $f_i$ removed from the dataset. We then calculated the stray light corrected CV image $u_i^\star = \convOp(h_i)^{-1} f_i$. Note that both $h_i$ and $u_i^\star$ are calculated without any assumption of a zero lunar disk in image $i$, so the values of $u_i^\star$ on the lunar disk represent an independent check on the correction's effectiveness. In Fig.~\ref{cvAnalysis} we compare the lunar disks of $f_i$ and $u_i^\star$ in 171~\AA. We find that the lunar disk values after deconvolution are very strongly clustered near zero, as seen in a histogram of the lunar disk values (\emph{bottom right}).

\begin{figure}
\includegraphics[scale=1.5]{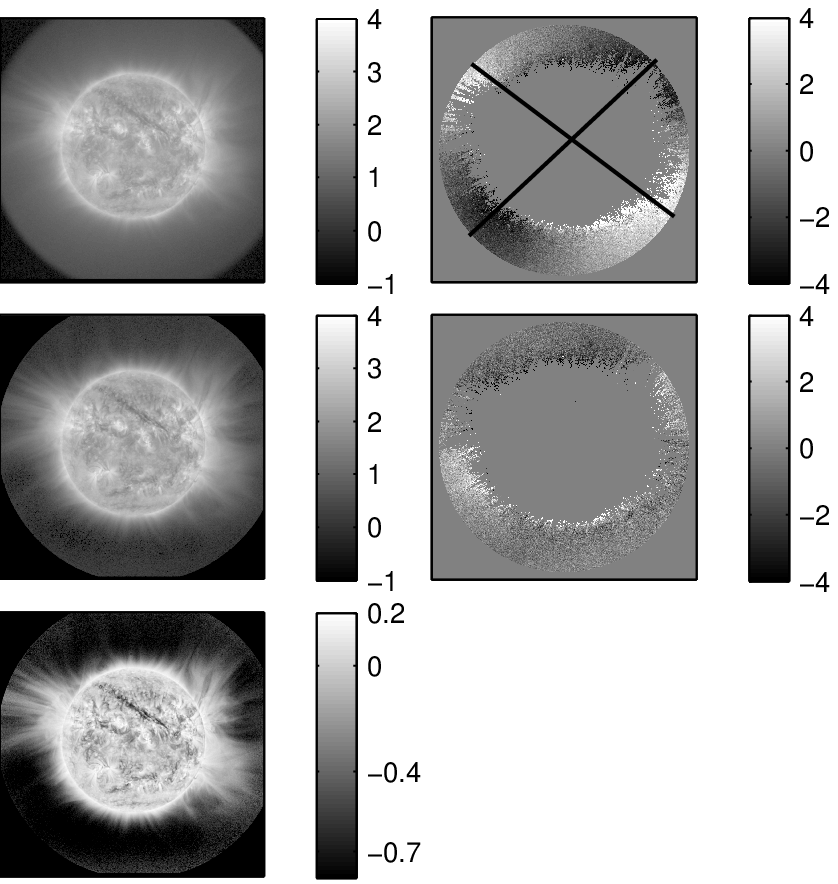}
\caption{Analysis of calibration roll images before and after stray light correction. \emph{Left:} The first roll image before stray light correction (\emph{top}), after correction (\emph{middle}), and fractional change ($(u_0 - f_0)/f_0$, \emph{bottom}). The colorbars for $f_0$ and $u_0$ are in units of $\log_{10} \, \text{DN}$. \emph{Right:} Sample difference image before correction ($\Delta f_{90}$, \emph{top}) and after ($\Delta u_{90}$, \emph{middle}), units of DN. The two black lines denote the axes of preferential scatter for $f_0$ and $f_{90}$.
}
\label{cal_roll}
\end{figure}

Calibration roll images provide direct evidence of anisotropic scatter and our deconvolution's ability to correct it. We will present the analysis for 171 \Ang here, leaving the other bands to future publications. On Nov.~8 2011, STEREO-B executed a $360^\circ$ calibration roll, and in each band, EUVI-B acquired 9 solar images at roll angles of 0, 60, 90, 120, 180, 240, 270, 300, and $360^\circ$ relative to the pre-roll position. These images are useful because the direction of anisotropic scatter rotates with STEREO-B, introducing discrepancies between the images. To pick out these discrepancies, an $8 \times 8$ boxcar was applied to reduce noise and the images were rotated into a common solar coordinate system. This procedure was repeated on stray light-corrected roll images, resulting in series $\{f_\rho\}_{\rho = 0}^{360}$ and $\{u_\rho\}_{\rho = 0}^{360}$ indexed by roll angle $\rho$. A haze of scattered light can be observed off the limb in the the image $f_{0}$ (Fig.~\ref{cal_roll}, \emph{top left}), which is much reduced in the corrected image $u_0$ (\emph{middle left}).

Anisotropic scatter rotation was tracked using the difference images $\Delta f_\rho = f_\rho - f_0$ and $\Delta u_\rho = u_\rho - u_0$. To avoid analyzing regions of the Sun that changed significantly over the course of the roll, we examined the difference $\Delta f_{360}$ between the pre- and post-roll images, and all pixels where $|\Delta f_{360}| > 1 \, \, \text{DN}$ were masked out of the difference images. In $\Delta f_{90}$ (\emph{top right}), we observe a `dark axis' and a $\approx90^\circ$-rotated `light axis' corresponding to the preferential scattering directions for $f_0$ and $f_{90}$ respectively. (These regions are very diffuse, so the separation angle can only be estimated up to about $\pm 10^\circ$.) These axes are eliminated in $\Delta u_{90}$ because the stray light correction greatly reduces the anisotropic haze. Similar reductions are observed at other $\rho$ values.

To estimate the error in the corrected images, we decompose the total error $\Err = u-u^{\text{true}}$ into components due to noise and PSF error, and estimate their contributions separately. To obtain the decomposition we substitute \eqrefa{imageFormModelEqn} into \eqrefa{uDef} and introduce the PSF error variable $\delta h~=~h~-~h^{\text{true}}$, obtaining
\begin{equation}
\Err = u-u^{\text{true}} =\convOp(h)^{-1} \convOp(\delta h) u^{\text{true}} + \convOp(h)^{-1} n.
\end{equation}		
The right hand side's first term gives the error due to PSF misestimation, which we name $\Err_{\text{psf}}$. The second, $\Err_{\text{noise}}$, is the noise in the corrected image.

To estimate $\Err_{\text{psf}}$, we first apply an $8 \times 8$ moving average filter to $u$ and $f$ to reduce noise, so that $u-u^{\text{true}} \approx \Err_{\text{psf}}$. (Averaging also reduces $\Err_{\text{psf}}$ on small scales, potentially leading to underestimation of the error. However, we are only interested in the large-scale effect of stray light, and on large scales, $\Err_{\text{psf}}$ is unaffected by averaging.) Next, we assume that $\Err_{\text{psf}}$ can be bounded by a quantity $\sigma_{\text{psf}}$ which is proportional to the magnitude of the stray light correction $|u - f|$:
\begin{equation} \label{boundSysError}
|u - u^{\text{true}}| \leq B|u - f| = \sigma_{\text{psf}}.
\end{equation}
We use the bounding term $B|u - f|$ because it shares two distinctive properties with $\Err_{\text{psf}}$: both are approximately invariant if $u^{\text{true}}$ is changed by scaling or addition of a constant. These invariance properties help ensure that an empirical bound based on lunar transit images alone will continue to be valid for other solar images. 

We determined the constant $B$ by requiring that \eqrefa{boundSysError} hold true over almost all of the lunar disk pixels of $f_i$ and $u_i^\star$. Setting $u = u_i^\star$, $f = f_i$, dividing both sides by $|u_i^\star - f_i|$, and noting that $u_i^{\text{true}} = 0$ on the lunar disk, we find
$|u_i^\star|/|u_i^\star - f_i| \leq B$ is required on the lunar disk pixels. We call the left-hand side ratio $b_i^\star$, and collect the values of $b_i^\star$ on the 8 lunar disks into a histogram. A normalized histogram for 171 \Ang is given in Fig.~\ref{cvErrorAnalysis}, \emph{right}, and values at the $68^{th}$, $95^{th}$, and $99.7^{th}$ percentiles, corresponding to the first three standard deviations of a Gaussian, are 0.08, 0.13, and 0.16. We set $B$ equal to the $95^{th}$ percentile of the histogram, which gives a conservative $2\sigma$ error bound robust to outliers. To estimate the contribution of noise to the error, we analytically calculate its covariance matrix using the known variance of $n$. The diagonal of this covariance matrix, $\sigma_{\text{noise}}^2$, is our estimate of the squared error due to noise. We add the two independent errors in quadrature to obtain the final error estimate $\sigma$:
%
$\sigma^2 = \sigma_{\text{psf}}^2 + \sigma_{\text{noise}}^2$.

\begin{figure}
\includegraphics[scale=2]{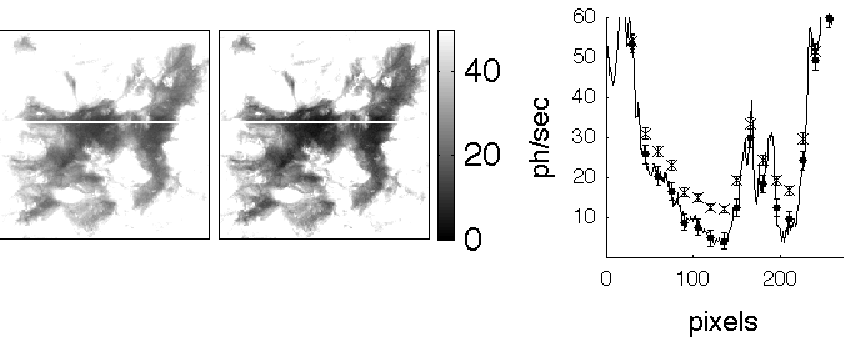}
\caption{Stray light correction of an on-disk coronal hole observed by EUVI-B on Nov.~$21$, 2008. \emph{Left and center:} The observed data $f$ and corrected data $u$ (ph/sec). \emph{Right:} Line plots of $f$ (crosses) and $u$ (solid line) along the white line segment. The horizontal axis is in pixels relative to the segment's left endpoint. Error bars are given every 20 pixels. \vspace{-1mm}}
\label{corHoleBeforeAfter}
\end{figure}

\section{RESULTS}

Here we note just a few of the effects of stray light correction on faint regions. Fig.~\ref{corHoleBeforeAfter} shows a coronal hole before and after correction. The stray light corrected coronal hole is significantly dimmer: the percent change $(u-f)/f$ ranges from $-40$ to $-70\%$ over most of the coronal hole, implying that 40-70\% of the apparent coronal hole emissions are stray light. The plot of intensity versus position (\emph{right}) contains the error bars $\sigma$. Note that they are small relative to the size of the correction, which is representative of typical results. Stray light correction also removes the haze seen above the limb (Fig.~\ref{cal_roll}, \emph{top and center left}), often reducing the intensity by 75\% or more (\emph{bottom left}). A filament just above the solar center is also reduced in intensity by 50-70\% after stray light correction (\emph{bottom left}).

These large downward corrections to faint regions have major impacts on the plasma diagnostics available from EUV images, which in turn are related to the electron temperature $T_e$\ and density $n_e$. Assuming electron impact excitation, the intensity of the emission in images from the $k^{th}$ spectral band is $I_k \propto \int_\mathrm{LOS} \mathrm{d} l \, n^2_e(l) Q_k(T_e(l))$, where $Q_k(T_e)$\ comes from an optically thin plasma emission model and $\int_\mathrm{LOS} \mathrm{d}l$\ represents integration along the line-of-sight \citep{Brown:1991}. This relationship may be utilized to create 3D tomographic maps of $n_e$ and $T_e$ \citep{Frazin09}. A downward correction of the observed intensity causes a proportional reduction of the estimated value of $n_e^2$, indeed, analysis of the 171, 195 and 284 \AA\ intensities show a downward revision of the coronal hole column density $[\int_\mathrm{LOS} \mathrm{d} l \, n^2_e(l) ]^{1/2}$ of $\sim 40\%$.  The effect on the estimated $T_e$ is more complex. The removal of stray light from the off-limb causes a dramatic steepening of the profile function $n_e(h) $\   (where $h$\ is the height above the photosphere).  The specific impact of stray light correction, including constraints on solar wind models from the corrected profiles $n_e(h), \, T_e(h)$, is currently under investigation.  Obvious consequences include reduction of the plasma $\beta$, electron-ion collision rates, and the mass of solar wind plasma requiring acceleration.
\section{CONCLUSION}
We have obtained PSFs for all 4 bands of STEREO-B/EUVI using lunar transit data, which enables us to correct all EUVI-B images for stray light and provide uncertainties.  Similar methods may be applied to treat the stray light problems in the other solar EUV imaging instruments (SOHO/EIT, TRACE, STEREO-A/EUVI, SDO/AIA). This work and its heliophysical implications will be reported in more detail in upcoming publications. 

\acknowledgments{We thank Jean-Pierre Wuelser for continuous assistance with EUVI technical issues; Frederic Auchere and Raymond Mercier, for useful discussions of PSF modeling; and Simon Plunkett, for granting our request for EUVI calibration roll data.}


\begin{thebibliography}{13}
\expandafter\ifx\csname natexlab\endcsname\relax\def\natexlab#1{#1}\fi

\bibitem[{Brown {et~al.}(1991)Brown, Dwivedi, {et~al.}}]{Brown:1991}
Brown, J.~C., Dwivedi, B.~N., {et~al.} 1991, Astronomy and Astrophysics, 249,
  277

\bibitem[{Cranmer(2010)}]{Cranmer2010}
Cranmer, S. 2010, Space Sci. Rev. \emph{in press.}

\bibitem[{DeForest {et~al.}(2009)DeForest, Martens, \&
  Wills-Davey}]{DeForest:CoronaStrayLightTRACE:2009}
DeForest, C.~E., Martens, P. C.~H., \& Wills-Davey, M.~J. 2009, The
  Astrophysical Journal, 690, 1264

\bibitem[{{Frazin} {et~al.}(2009){Frazin}, {V{\'a}squez}, \&
  {Kamalabadi}}]{Frazin09}
{Frazin}, R.~A., {V{\'a}squez}, A.~M., \& {Kamalabadi}, F. 2009, The
  Astrophysical Journal, 701, 547

\bibitem[{Gburek {et~al.}(2006)Gburek, Sylwester, \&
  Martens}]{Gburek:TracePSF171:2006}
Gburek, S., Sylwester, J., \& Martens, P. 2006, Solar Physics, 239, 531

\bibitem[{Golub \& Pereyra(2002)}]{Golub:separablenonlinear:2002}
Golub, G. \& Pereyra, V. 2002, in Institute of Physics, Inverse Problems, 1--26

\bibitem[{Goodman(1996)}]{Goodman:FourierOptics}
Goodman, J.~W. 1996, Introduction to Fourier Optics, 2nd edn. (McGraw-Hill)

\bibitem[{Howard {et~al.}(2008)}]{Howard08}
Howard, R.~A. {et~al.} 2008, Space Sci. Rev., 136, 67

\bibitem[{Krautschik {et~al.}(2002)Krautschik, Ito, Nishiyama, \&
  Okazaki}]{Krautschik:EUVscatter:2002}
Krautschik, C.~G., Ito, M., Nishiyama, I., \& Okazaki, S. 2002, Proc. SPIE,
  4688, 289

\bibitem[{Krieger {et~al.}(1973)Krieger, Timothy, \& Roelof}]{Krieger1973}
Krieger, A.~S., Timothy, A.~F., \& Roelof, E.~C. 1973, Solar Physics, 29, 505

\bibitem[{Mart{\'\i}nez-Galarce {et~al.}(2010)Mart{\'\i}nez-Galarce, Harvey,
  {et~al.}}]{Galarce:EUVModeling:2010}
Mart{\'\i}nez-Galarce, D., Harvey, J., {et~al.} 2010, Proc. SPIE, 7732, 773237

\bibitem[{Murtagh {et~al.}(1995)Murtagh, Starck, \&
  Bijaoui}]{MurtaghStarck:ImRestVST:1995}
Murtagh, F., Starck, J.-L., \& Bijaoui, A. 1995, Astronomy and Astrophysics,
  112, 179

\bibitem[{Starck {et~al.}(2002)Starck, Pantin, \& Murtagh}]{Starck02}
Starck, J.~L., Pantin, E., \& Murtagh, F. 2002, PASP, 114, 1051

\end{thebibliography}
\end{document}